\documentclass[runningheads]{llncs}
\usepackage[T1]{fontenc}
\usepackage{graphicx,verbatim}
\usepackage{cases}
\usepackage{xfrac}
\usepackage{float}
\makeatletter
\@ifundefined{endsubequations}{}{%

}
\makeatother
\usepackage{amsmath}
\usepackage{amssymb}
\usepackage{multirow, tabularx}
\usepackage{algorithm}
\usepackage{algpseudocode}
\newcommand{\BlankLine}{\Statex}
\newcommand{\COMMENT}[1]{\Comment{#1}}
\usepackage{enumitem}
\usepackage{pifont}
\usepackage{xcolor}
\definecolor{cvprblue}{rgb}{0.21,0.49,0.74}
\usepackage[pagebackref,breaklinks,colorlinks,citecolor=cvprblue]{hyperref}

\newcolumntype{C}{>{\centering\arraybackslash}X}

\DeclareUnicodeCharacter{2009}{\,}
\begin{document}
\title{EcoScale‑Net: A Lightweight Multi‑Kernel Network for Long‑Sequence 12-lead ECG Classification}
%
% \begin{comment}  %% Removed for anonymized MICCAI 2025 submission
\author{Dong-Hyeon Kang\inst{1} \and
Ju-Hyeon Nam\inst{1} \and
Sang-Chul Lee\inst{1, 2}}
\institute{Department of Electrical and Computer Engineering, Inha University, Republic of Korea \and
DeepCardio, Republic of Korea \\
\email{\{orionis2001, jhnam0514\}@inha.edu} \email{sclee@\{inha.ac.kr, deepcardio.com\}}\\}
% \end{comment}

% \author{Anonymized Authors}  %% Added for anonymized MICCAI 2025 submission
% \authorrunning{Anonymized Author et al.}
% \institute{Anonymized Affiliations \\
%     \email{email@anonymized.com}}

\maketitle              % typeset the header of the contribution
\begin{abstract}
Accurate interpretation of 12‑lead electrocardiograms (ECGs) is critical for early detection of cardiac abnormalities, yet manual reading is error‑prone and existing CNN‑based classifiers struggle to choose receptive‑field sizes that generalize to the long sequences typical of ECGs. Omni‑Scale CNN (OS‑CNN) addresses this by enumerating prime‑sized kernels inspired by Goldbach’s conjecture to cover every scale, but its exhaustive design explodes computational cost and blocks deeper, wider models. We present \textit{\textbf{\underline{E}fficient \underline{C}onvolutional \underline{O}mni-\underline{Scale} \underline{Net}work (\underline{EcoScale-Net})}}, a hierarchical variant that retains full receptive‑field coverage while eliminating redundancy. At each stage, the maximum kernel length is capped to the scale still required after down‑sampling, and $1 \times 1$ bottleneck convolutions inserted before and after every Omni‑Scale block curtail channel growth and fuse multi‑scale features. On the large‑scale CODE‑15\% ECG dataset, EcoScale-Net reduces parameters by 90\% and FLOPs by 99\% compared with OS‑CNN, while raising macro‑averaged F1‑score by 2.4\%. These results demonstrate that EcoScale-Net delivers state‑of‑the‑art accuracy for long‑sequence ECG classification at a fraction of the computational cost, enabling real‑time deployment on commodity hardware. Our EcoScale-Net code is available in \href{https://github.com/Inha-CVAI/EcoScaleNet_MICCAIW2025}{GitHub Link}.
% Authors must provide keywords and are not allowed to remove this Keyword section.

\end{abstract}

\section{Introduction}
Cardiovascular disease is the leading global cause of death, and 12‑lead electrocardiograms (ECGs) remain the most widely used non‑invasive modality for detecting cardiac abnormalities. Manual interpretation, however, is error‑prone owing to noise and inter‑observer variability \cite{chung2022clinical}. These constraints have accelerated interest in automated ECG analysis based on deep learning, a trend already validated in other medical‑vision domains \cite{nam2024modality}.

Early ECG classifiers combined convolutional neural networks (CNNs) with recurrent units to exploit both local morphology and long‑range rhythm cues, achieving super‑human accuracy in some tasks \cite{tan2018application}. Subsequent work expanded to multi‑lead, multi‑label settings \cite{chen2019large} and even image‑based ECG interpretation \cite{nam2024vizecgnet}. Yet these architectures still depend on manually tuned kernel sizes or dilations, which scale poorly to the long sequences characteristic of ECG signals.

Omni‑Scale CNN (OS‑CNN) \cite{tang2022omniscale} sidesteps manual tuning by enumerating prime‑sized kernels—via Goldbach’s conjecture—to cover every receptive‑field length. Unfortunately, this exhaustive design inflates parameters and FLOPs quadratically with input length, throttling model depth and width; follow‑ups such as OSGAN \cite{chen2024osgan} and TF‑Net \cite{lei2024time} inherit the same bottleneck.

To address these issues, we introduce \textit{\textbf{\underline{E}fficient \underline{C}onvolutional \underline{O}mni-\underline{Scale} \underline{Net}work (\underline{EcoScale-Net})}}, a hierarchical reformulation of OS‑CNN that preserves full receptive‑field coverage while eliminating redundancy. This straightforward architectural modification enables the proposed model to leverage Omni-Scale Convolution within a hierarchical framework, progressively reducing the required coverage range at deeper layers. This modification facilitates more comprehensive feature extraction and allows for a deeper and wider design than the OS-CNN. Evaluated on the large‑scale CODE‑15\% dataset \cite{ribeiro2021code}, EcoScale‑Net cuts parameters by 90\% and FLOPs by 99\% relative to OS‑CNN, while boosting macro‑averaged F1‑score by 2.4\%—the best among all compared models. The contributions of this study are as follows:

\begin{itemize}
\item We introduce a novel structure (\textit{\textbf{EcoScale-Net}}) aimed at efficiently configuring the kernel size for processing long signals such as ECG. Compared to OS-CNN, our method allows for a deeper and wider design using a hierarchical framework, leading to a reduction in parameters by over tenfold and FLOPs by more than 99\%, while achieving a 2.4\% performance enhancement. 

\item We demonstrate that applying convolution with a kernel size of 1 before and after each OS stage effectively prevents an explosive increase in channel dimensions while modeling the relationships of each receptive field. 

\item Despite having lower FLOPs and fewer parameters compared to most other models, it demonstrates the best performance in terms of the average F1-score among the various models we compared. 
\end{itemize}

\section{Method}

\subsection{Preliminary: Omni‑Scale Convolution}

Omni‑Scale Convolution (OS‑Conv) was introduced to avoid \emph{manual} kernel‑size tuning in long signals.  
Its key idea is to enumerate prime–sized kernels so that every possible receptive‑field (RF) length can be obtained by \emph{adding} three 1‑D convolutions.

\paragraph{Kernel set.}
Let $\mathbb{P}^{(j)}$ be the kernel set at OS stage $j$ ($j=1,2,3$):
\begin{equation}
\label{eq:kernel_set}
\mathbb{P}^{(j)}=
\begin{cases}
\{\,p\mid p\ \text{is prime},\;1\le p\le p_k\} & (j=1,2),\\[2pt]
\{1,2\} & (j=3),
\end{cases}
\end{equation}
where $p_k$ is the \emph{largest} prime allowed at the first two stages.

\paragraph{Receptive‑field coverage.}
For $p^{(j)}\!\in\!\mathbb{P}^{(j)}$, the RF length produced by the three‑stage stack is
\begin{equation}
\label{eq:rf_length}
S = p^{(1)}+p^{(2)}+p^{(3)}-2 .
\end{equation}
Because any even number $e$ $(\le 2p_k)$ can be written as the sum of two primes ($p^{(1)}{+}p^{(2)}$) by Goldbach’s conjecture,~\eqref{eq:rf_length} yields
\[
\mathbb{S}= \bigl\{e,\;e-1 \;\big|\; e\in\mathbb{E} \bigr\},\qquad
\mathbb{E}=\{2,4,\dots,2p_k\},
\]
so \emph{all} even and adjacent odd RF lengths up to $2p_k$ are covered.

\paragraph{Computational bottleneck.}
Enumerating every prime up to $p_k$ incurs
$O\!\bigl(\sum_{p\le p_k}p\bigr)=O(p_k^2)$ parameters and FLOPs per stage; thus complexity grows \emph{quadratically} with $p_k$ and linearly with sequence length. For long ECG sequences ($\sim$4 k samples) this leads to impractical memory and time costs, preventing deeper or wider models.  The remainder of this paper presents a hierarchical reformulation that keeps full RF coverage while eliminating this redundancy.

\begin{figure*}[t]
    \centering
    \includegraphics[width=\textwidth]{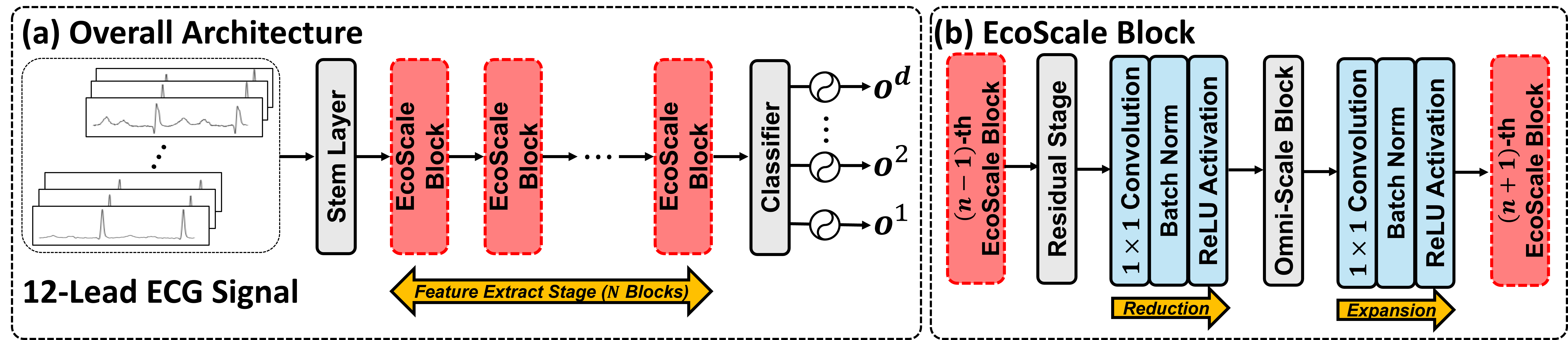}
    \caption{(a) Overview of EcoScale‑Net. (b) EcoScale Block which is main module for proposed approach.}
    \label{fig:EcoScaleNet}
\end{figure*}

\subsection{Overall Architecture}
Our main goal is to \textit{efficiently configure the kernel size to cover the receptive field for signals with arbitrary lengths, leveraging the key advantage of OS convolution}. To achieve main goals and address such impractical issues for long-range 12-ECG, we propose the \textit{\textbf{\underline{E}fficient \underline{C}onvolutional \underline{O}mni-\underline{Scale} \underline{Net}work (\underline{EcoScale-Net})}}, a lightweight reformulation of OS‑CNN for multi‑label ECG classification (Fig.~\ref{fig:EcoScaleNet}). EcoScale‑Net replaces OS‑CNN’s exhaustive kernel list with a \emph{hierarchical} design and dual \(1\times1\) bottlenecks, preserving full receptive‑field (RF) coverage while cutting parameters and FLOPs. Note that we select same kernel selection strategy with OS-CNN (Algorithm \ref{alg:prime_set_calculation}).  In this paper, we used ResNet-34 \cite{he2016deep} as the backbone for the hierarchical architecture.

Let $\mathbf{X} = \{ x_{1}, x_{2}, \dots, x_{12} \} \in \mathbb{R}^{12 \times T}$ be the input 12-lead ECG signals where $T$ is the length of the ECG signals. Then, we extract the fundamental features $z_{0}$ from the input ECG signal $\mathbf{X}$ as follows:
\begin{equation}
    z_0 = s(\mathbf{X})\in \mathbb{R}^{C_0 \times L_0}
\end{equation}
\noindent where $s ( \cdot )$ denotes the stem layer, which consists of a convolution with a kernel size of 7 and a stride of 2, followed by batch normalization, ReLU, and max pooling for downsampling. And, $C_{0}$ and $L_{0}$ are the numbers of channels and time length of the stem feature $z_0$, respectively. Subsequently, the stem feature $z_{0}$ is processed through a feature extractor composed of $N$ stages. At $i$-th feature extractor stage, we can represent feature $z_{i}$ as follows:
\begin{equation}
    \begin{cases}
        &h_i = r_{i}(z_{i-1})\in{\mathbb{R}^{C_i \times L_i}} \\
        &z_i = e_{i}(h_i)\in{\mathbb{R}^{C_i \times L_i}}
    \end{cases}
\end{equation}
\noindent where $r_{i} (\cdot)$ and $e_{i} (\cdot)$ are the residual stage and EcoScale block, respectively. For convenience, we let $f_{i} = e_{i} \circ r_{i}$. And, $C_{i}$ and $L_{i}$ denote the number of channels and time length at $i$-th feature extractor stage, respectively. The feature $z_{N} \in \mathbb{R}^{C_{N} \times L_{N}}$, which is processed through the final $N$-th stage, is then fed into a classifier $g (\cdot)$ comprising Global Average Pooling (GAP) and a fully connected layer to produce the final prediction $p = g(z_{N}) \in \mathbb{R}^{M}$ where $M$ is the number of classes contained in datasets. This overall process can be formally described as follows:
\begin{equation}
\underbrace{p}_\text{Final Prediction} =
(\underbrace{g}_\text{Classifier} \circ
\underbrace{\mathop{\bigcirc}\limits_{i = 1}^{N}\ f_{i}}_\text{Feature Extractor with $N$ stages} \circ
\underbrace{s}_\text{Stem Layer}) (\mathbf{X})
\end{equation}
\noindent where $\mathop{\bigcirc}\limits_{i = 1}^{N} f_{i} = f_{N} \circ f_{N - 1} \circ \cdots \circ f_{2} \circ f_{1}$. And, we use the binary cross entropy (BCE) loss $\mathcal{L}_{\text{bce}}$ to optimize the parameters of stem layer $s$, feature extractor $f_{i}$ for $i = 1, \dots, N$, and classifier $g$ as follows:
\begin{equation}
    \mathcal{L} (p, t) = \sum_{m = 1}^{M} \mathcal{L}_{BCE} (p^{m}, t^{m})
\end{equation}
\noindent where $t$ denotes the ground truth label corresponding to $p$. Hence, $t^{m}$ is $m$-th class label.

\begin{algorithm}[H]
\caption{Selection of kernel sizes}
\label{alg:prime_set_calculation}
\begin{algorithmic}[1]
\Require length to cover $L$
\Ensure Set of kernel sizes $\{1, 2, \ldots, p_k\}$

\BlankLine
\State $p_k \gets \text{smallest prime satisfying } 2 \cdot p_k > L$
\COMMENT{Find the prime such that $2 \cdot p_k$ exceeds the cover length $L$}

\BlankLine
\State $\text{Prime Set} \gets \{\,p \le p_k \mid p \text{ is prime}\,\}$
\COMMENT{Collect all primes up to $p_k$}

\BlankLine
\State \Return $\text{Prime Set}$
\end{algorithmic}
\end{algorithm}

\subsection{EcoScale Residual Block}

\paragraph{Hierarchical RF cascading.}
Stage $i$ covers an effective signal length
\(l_i=L_{i-1}/d_i\) (\(d_i\) = cumulative down‑sampling factor).
We choose the smallest prime \(p_k^{(i)}\) satisfying
\(2p_k^{(i)}>l_i\) and set the stage kernel set
\(
\mathbb{P}^{(i)}=\{p\le p_k^{(i)}\mid p \text{ is prime}\}\cup\{1,2\}.
\)
Thus deeper stages use \emph{smaller} maximum kernels, avoiding
redundant large filters while retaining complete RF coverage.

\paragraph{Channel‑efficient fusion.}
Each block applies
\[
\text{conv}_{1\times1}\!\rightarrow\!\text{(parallel OS‑Conv)}\!\rightarrow\!
\text{concat}\!\rightarrow\!\text{conv}_{1\times1},
\]
where the first \(1\times1\) bottleneck reduces channels to $C_i/2$
and the second restores them, preventing the quadratic channel
explosion observed in vanilla OS‑CNN.

\subsection{Complexity Analysis and Receptive Field of EcoScale-Net}
While EcoScale-Net enhances computational efficiency by hierarchically reducing the length to be covered, we conducted an analysis to demonstrate its efficiency compared to OS-CNN, even when maintaining the same feature length. Using the prime number theory ($\sum_{x \in \{ 1, 2, 3, 5, 7, \dots, p_{k} \}} x \approx \frac{p^{2}_{k}}{\ln (p_{k})}$), we can theoretically compare the parameter counts of OS-CNN and EcoScale-Net. For convenience, we consider only the first stage and the number of channels of the input feature signal is $C_{i}$. Then, the ratio of number of parameters between OS-CNN and EcoScale-Net can be represented as follows:
\begin{equation}
\label{eq:parameter_analysis}
    \begin{split}
    &\frac{(1 \times C_{i} \times \frac{C_{i}}{2}) + \sum_{x \in \{ 1, 2, 3, 5, 7, \dots, p_{k} \}} (x \times \frac{C_{i}}{2} \times \frac{C_{i}}{2})}{\sum_{x \in \{ 1, 2, 3, 5, 7, \dots, p_{k} \}} (x \times C_{i} \times C_{i})} \\
    &= \frac{\frac{C^{2}_{i}}{2} + \frac{C^{2}_{i}}{4} \sum_{x \in \{ 1, 2, 3, 5, 7, \dots, p_{k} \}} x}{C^{2}_{i} \sum_{x \in \{ 1, 2, 3, 5, 7, \dots, p_{k} \}} x} \\
    &= \frac{\frac{1}{2} + \frac{1}{4} \sum_{x \in \{ 1, 2, 3, 5, 7, \dots, p_{k} \}} x}{\sum_{x \in \{ 1, 2, 3, 5, 7, \dots, p_{k} \}} x} \\ &\approx \frac{\frac{1}{2} + \frac{1}{4} \cdot \frac{p^{2}_{k}}{\ln (p_{k})}}{\frac{p^{2}_{k}}{\ln (p_{k})}} \\
    &= \frac{2\ln (p_{k}) + p^{2}_{k}}{4p^{2}_{k}}
    \end{split}
\end{equation}

With \(p_k\!=\!11\) (CODE‑15\%), EcoScale‑Net requires only \(\approx\!26\%\) of the first‑stage parameters of OS‑CNN, and the gap widens in deeper layers due to the hierarchical reduction of \(p_k^{(i)}\).

A $1 \times 1$ convolution recombines channels at the same timestamp and therefore contributes zero additional span to the receptive field (RF). For a sequence of kernels with lengths $\{l_i\}_{i=1}^n$ (stride 1, no dilation) the cumulative RF is $R = \sum_{i = 1}^{n} l_{i} - (n - 1)$. Hence, the dual $1 \times 1$ bottlenecks that frame every multi-kernel block in EcoScale-Net neither shrink nor reset temporal reach, while still enabling channel‑wise projections and fusions.

\begin{table*}
    \centering
    \scriptsize
    \setlength\tabcolsep{1.5pt}

    \begin{tabular}{c|c|c|ccc|ccc|ccc}
    \hline
    \multicolumn{1}{c|}{\multirow{2}{*}{Model}} & \multicolumn{1}{c|}{\multirow{2}{*}{Param (M)}} & \multicolumn{1}{c|}{\multirow{2}{*}{FLOPs (G)}} & \multicolumn{3}{c|}{\textbf{Average}} & \multicolumn{3}{c|}{\textbf{Exp‑\textsc{ML}}} & \multicolumn{3}{c}{\textbf{Exp‑\textsc{BIN}}} \\ \cline{4-12}
     & & & Pre & Rec & F1 & Pre & Rec & F1 & Pre & Rec & F1 \\
     \hline
      GRU \cite{69e088c8129341ac89810907fe6b1bfe} & 0.55 & 145.5 & 89.7 & 84.0 & 86.0 & 89.1 & 77.9 & 81.9 & 90.3 & 90.0 & 90.1  \\

      Transformer \cite{NIPS2017_3f5ee243} & 87.43 & 64.1 & 88.8 & 86.9 & 87.7 & 85.7 & 81.6 & 83.5 & 91.8 & 92.2 & 91.9  \\ 

      ResNet \cite{he2016deep} & 3.85 & 45.9 & 92.4 & 93.2 & 92.8 & 91.8 & 93.2 & 92.5 & 92.9 & 93.1 & 93.0  \\ 
      
      InceptionTime \cite{ismail2020inceptiontime} & 0.49 & 129.4 & - & - & - & 92.8 & 89.5 & 91.1 & - & - & - \\

      ResBlk \cite{ribeiro2020automatic} & 6.78 & 124.4 & \textcolor{blue}{\textbf{\textit{93.6}}} & \textcolor{blue}{\textbf{\textit{94.0}}} & \textcolor{blue}{\textbf{\textit{93.8}}} & 92.8 & \textcolor{blue}{\textbf{\textit{93.8}}} & 93.2 & \textcolor{blue}{\textbf{\textit{94.4}}} & \textcolor{blue}{\textbf{\textit{94.1}}} & \textcolor{blue}{\textbf{\textit{94.3}}} \\

      OS-CNN \cite{tang2022omniscale} & 84.83 & 22240.2 & - & - & - & 91.6 & 91.4 & 91.5 & - & - & -  \\

      ResU-Dense \cite{10404234} & 7.20 & 278.6 & 92.8 & 93.7 & 93.2 & 91.8 & 93.6 & 92.6 & 93.8 & 93.8 & 93.8 \\ 

      ResUNet-LC \cite{hwang2024multi} & 9.98 & 3749.1 & 81.8 & 81.9 & 81.8 & \textcolor{red}{\textbf{\underline{94.1}}} & 92.7 & \textcolor{blue}{\textbf{\textit{93.3}}} & 69.5 & 71.0 & 70.3 \\ 

      \hline
      \multicolumn{1}{c|}{\multirow{2}{*}{\textbf{EcoScale-Net}}} & \multicolumn{1}{c|}{\multirow{2}{*}{8.55}} & \multicolumn{1}{c|}{\multirow{2}{*}{66.9}}
      & \textcolor{red}{\textbf{\underline{94.4}}} & \textcolor{red}{\textbf{\underline{94.4}}} & \textcolor{red}{\textbf{\underline{94.3}}} & \textcolor{blue}{\textbf{\textit{93.6}}} & \textcolor{red}{\textbf{\underline{94.2}}} & \textcolor{red}{\textbf{\underline{93.9}}} & \textcolor{red}{\textbf{\underline{95.1}}} & \textcolor{red}{\textbf{\underline{94.5}}} & \textcolor{red}{\textbf{\underline{94.7}}} \\
      & & & \textbf{+0.5} & \textbf{+0.4} & \textbf{+0.5} & \textbf{-0.5} & \textbf{+0.4} & \textbf{+0.6} & \textbf{+0.7} & \textbf{+0.4} & \textbf{+0.4} \\

    \hline
    \end{tabular}
    \caption{\label{tab:exp_res1}Experiment results on the ECG dataset. We also provide number of trainable parameters (M) and FLOPs (G) for each methods.} \vspace{-1.5cm}
\end{table*}

\section{Experimental Results}
\subsection{Experimental Settings and Implementation Details}
We conducted all experiments in Pytorch 2.1.2 and Python 3.11.5 on a NVIDIA GeForce RTX 4070 GPU and used a publicly available ECG dataset \cite{ribeiro2021code} for training and evaluation. Each record is represented as a $12\times4096$ signal; the characteristic period of $\sim256$ samples guided the initial maximum prime kernel size $p_k$ in our architecture. To evaluate EcoScale‑Net under different clinical use‑cases, we defined \emph{two} tasks, both using identical patient‑level splits (90\% train, 5\% validation, and 5\% test, stratified by patient~ID to avoid leakage):

\begin{enumerate}[label=(\arabic*), leftmargin=10pt]
    \item \textbf{Positive‑Only Multi‑Label Classification (Exp‑\textsc{ML}).}  
    We retained only \textbf{positive} ECGs that contain at least one of the six target abnormalities \{1dAVb, RBBB, LBBB, SB, AF, ST\}.  
    The goal is to predict the \emph{subset} of abnormal labels present in each record.  
    This protocol emphasizes detailed abnormality recognition while forcing the network to focus on pathological patterns without the overwhelming prevalence of normal signals.

    \item \textbf{Binary Normal\,/\,Abnormal Classification (Exp‑\textsc{BIN}).}  
    The full cohort---both normal (negative) and abnormal (positive) recordings---was used.  
    A single binary label indicates whether \emph{any} of the six disorders occurs.  
    This setting reflects a real‑world screening scenario in which the foremost question is “\emph{Is this ECG normal?}”
\end{enumerate}

We compared the proposed EcoScale‑Net with five general signal classification models (GRU \cite{69e088c8129341ac89810907fe6b1bfe}, Transformer \cite{NIPS2017_3f5ee243}, ResNet \cite{he2016deep}, InceptionTime \cite{ismail2020inceptiontime}, and OS-CNN \cite{tang2022omniscale}) and three ECG signal abnormal detection models (ResBlk \cite{ribeiro2021code}, ResU-Dense \cite{10404234},  and ResUNet-LC \cite{hwang2024multi}).  Specifically, OS-CNN stacked three OS blocks with residual connections. All models were trained in an end-to-end manner using the AdamW \cite{loshchilov2017decoupled} optimizer. The initial learning rate started from $10^{-4}$ and was decreased to $10^{-6}$ using the cosine annealing learning rate scheduler \cite{loshchilov2016sgdr}, and the training settings were set to a batch size of 64 and epochs of 50. For evaluation, we used three metrics (Precision, Recall, and F1-Score) to measure the performance of each model. In all tables, \textcolor{red}{\textbf{\underline{Red}}} and \textcolor{blue}{\textbf{\textit{Blue}}} are the first and second best performance results, respectively.

\begin{table}
    \vspace{-0.0cm}
    \centering
    \scriptsize
    \setlength\tabcolsep{11pt}
        \begin{tabular}{c|c|c}
        \hline
            \centering
            
            \textbf{Model} & \textbf{Win} & \textbf{Average Rank} \\
            \hline
            GRU & 1 & 7.9 \\
            Transformer & 0 & 8.4 \\
            ResNet & 1 & 4.7 \\
            InceptionTime & 1 & 5.2 \\
            OS-CNN & 1 & 4.9 \\
            ResBlk & 2 & 3.3 \\
            ResU-Dense & 1 & 4.5 \\
            ResUNet-LC & 4 & 3.1 \\
            \hline
            \textbf{EcoScale-Net (Ours)} & \textbf{8} & \textbf{2.2} \\
            \hline
        \end{tabular} 
    \caption{\label{tab:exp_res2} Win counts and average ranks for Precision, Recall, and F1 score for each label.} \vspace{-1.5cm}
\end{table}

\subsection{Results Analysis}
Table \ref{tab:exp_res1} compares eight models with the proposed EcoScale‑Net on two complementary tasks derived from the same CODE‑15\% dataset: (i) a positive‑only multi‑label setting \textbf{(Exp‑ML}) and (ii) a binary normal/abnormal setting that includes both negative and positive records (\textbf{Exp‑BIN}). 

EcoScale‑Net achieves the best performance in both experimental configurations. In the positive‑only multi‑label task (Exp‑ML) it achieves a F1-score of 93.9\%, surpassing the next‑strongest CNN baseline (ResUNet-LC) by 0.6\% and the computationally intensive OS‑CNN by 2.4\%. When the problem is reduced to binary normal/abnormal screening (Exp‑BIN), it again leads with 94.7\% F1-score, maintaining a consistent margin over all eight models. These outcomes verify that the proposed hierarchical receptive‑field cascade and dual $1 \times 1$ bottleneck scheme capture both local morphology and global rhythm cues more effectively than conventional single‑scale or exhaustive multi‑scale alternatives. 

The accuracy gains do not come at the expense of efficiency. EcoScale‑Net contains 8.6M parameters and 66.9GFLOPs, yielding a 90\% parameter reduction and a 99\% FLOP reduction relative to OS‑CNN, while out‑performing lightweight architectures such as InceptionTime by 2.8\% in macro‑F1. Moreover, it is at least 50$\times$ leaner than U‑shaped models (e.g., ResUNet‑LC) that still fall short in accuracy, thereby establishing the state-of-the-art model for ECG analysis and enabling real‑time inference on commodity GPUs and edge devices. 

Robustness across diagnostic categories on Exp‑ML is evidenced by the win‑count statistics in Table \ref{tab:exp_res2}: EcoScale‑Net secures eight per‑label victories and the highest mean rank (2.2) for precision, recall, and F1—double the win total of its closest competitor. This uniform superiority indicates that the architecture generalises beyond a single arrhythmia class, delivering clinically reliable predictions across heterogeneous cardiac conditions while satisfying stringent computational budgets.

\begin{table}
    \centering
    \scriptsize
    \setlength\tabcolsep{4.5pt}

    \begin{minipage}{\columnwidth}
            \begin{tabular}{c|cc|c|cc}
            \hline
            \textbf{Model} & \textbf{OS Conv} & \textbf{C1D (1)} &  \textbf{Performance} & \textbf{Param (M)} & \textbf{FLOPs (G)} \\ 
            \hline
            Backbone & $\times$ & $\times$ & 92.5 & 3.85 & 45.9 \\
            EcoScale-Net* & $\checkmark$ & $\times$ &  93.2 & 9.70 & 167.7 \\
            \textbf{EcoScale-Net} &$\checkmark$ & $\checkmark$  & \textbf{93.9} & 8.55 & 66.9 \\ 
            \hline
        \end{tabular}
    \end{minipage}
    \caption{\label{tab:ablation}Ablation study for EcoScale-Net. C1D (1) and '*' denote a convolution 1D with kernel size 1 and the model without applying C1D (1), respectively.}
\end{table}

\subsection{Ablation Study}
In this section, we conducted an ablation study on the abnormal signal classification for 12-lead ECG signals to demonstrate the effectiveness of the proposed module. As listed in Table \ref{tab:ablation}, our approach, which utilized both OS Convolution (OS Conv) and convolution with kernel size 1 (C1D (1)) exhibited the best performance with reasonable computational complexity. We initially observed that the EcoScale-Net without C1D (1), achieved a 0.7\% performance improvement compared to the backbone, but with significantly higher complexity, including a 60\% and 72\% increase in the number of parameters and FLOPs, respectively. Conversely, applying C1D (1) at each OS stage reduced the number of parameters by 1.2M and decreased FLOPs by 100G while achieving a 0.7\% performance improvement. These results indicate that C1D (1) effectively models the dependencies of each receptive field, while mitigating the exponential increase in dimensions caused by channel concatenation.

\section{Conclusion and Future Works}
We have introduced EcoScale‑Net, a hierarchical omni‑scale convolutional network that attains complete receptive‑field coverage for long 12‑lead ECG sequences while curbing the quadratic cost growth of prior OS‑CNNs through stage‑wise kernel capping and dual convolutional bottlenecks. On the CODE‑15\% benchmark, EcoScale‑Net delivered state‑of‑the‑art macro‑F1 scores—93.9\% for positive‑only multi‑label diagnosis and 94.7\% for binary screening—using 90\% fewer parameters and 99\% fewer floating‑point operations than the original OS‑CNN, thereby establishing a new accuracy–efficiency Pareto frontier suitable for real‑time, low‑power clinical deployment.

Nevertheless, EcoScale‑Net still faces several limitations that we are actively working to overcome. First, we will conduct comprehensive backbone‑swap ablations and benchmark the model against additional lightweight architectures—such as MobileNet‑v4‑TS and MiniRocket—to verify the generality of our design. Second, we will validate the network on diverse public ECG datasets (PTB‑XL, Chapman‑SHA, and Shaoxing) to assess robustness across patient demographics, noise profiles, and recording hardware. In parallel, future work will extend the omni‑scale paradigm to other biosignals, incorporate adaptive kernel selection, and enable on‑device continual learning, thereby expanding the translational impact of EcoScale‑Net in both clinical and resource‑constrained environments.

\section*{Acknowledgements}

This work was supported in part by Institute of Information and communications Technology Planning \& Evaluation (IITP) grant funded by the Korea government (MSIT) (No.RS-2022-00155915, Artificial Intelligence Convergence Innovation Human Resources Development (Inha University).

\section*{Disclosure of Interests}
The authors have no competing interests to declare that are relevant to the content of this article.

\bibliographystyle{IEEEbib}
\bibliography{main}





% LNCS Homepage, \url{http://www.springer.com/lncs}, last accessed 2023/10/25
% \end{thebibliography}
\end{document}